\documentclass{ws-p8-50x6-00}

\begin{document}

\title{STAR results from the first RHIC year}

\author{Rene Bellwied for the STAR Collaboration}

\address{Physics Department, Wayne State University,
         Detroit, MI 48201, U.S.A.}


\maketitle

\abstracts{
An overview is presented of the results of the STAR experiment 
from the analysis of Au+Au collisions at $\sqrt{s_{_{NN}}} = 130$~GeV
acquired during the first year of RHIC operation. 
The transverse momentum distribution of negative hadrons has been studied 
out to 6 GeV/$c$ in central Au+Au collisions and elliptic flow measurements 
out to 4.5 GeV/$c$ in non-central collisions. This allows us to study
hard processes and potentially effects of nuclear matter on parton
propagation. Soft processes, which dominate particle production below 
around 2 GeV/$c$, have also been studied through the identified spectra of 
kaons, $\overline{p}$, and strange baryons.
}

\section{The STAR experiment and its physics goals}

The recent announcement of indirect evidence for the formation
of a new state of matter in several CERN heavy-ion experiments has led 
to renewed interest for the field of relativistic heavy-ion 
collisions. 
In its first run, the Relativistic Heavy Ion Collider
(RHIC) at the Brookhaven National Laboratory collided Au ions at a center
of mass energy of 130 GeV per nucleon pair, some 7.5 times higher than 
that previously possible in fixed target experiments at CERN. The ultimate
goal of experiments is to establish the existence of a deconfined phase of
quarks and gluons and to study the behaviour of strongly interacting matter 
under these conditions. 
As expected, the very first measurements reported from RHIC were based on 
hadronic observables. Thermal model descriptions of hadronic particle
production at CERN, in particular the strange particle abundances and 
ratios, have concluded that only the inclusion of a phase transition 
into any model calculation will allow a proper description of  the measured 
results~\cite{heinz}. In addition, particle production at high transverse 
momentum ($p_T$) is expected to become more pronounced at RHIC energies. 
High $p_T$ particles  originate from hard processes occuring during the 
earliest stages of the  collision and may therefore offer new insight into
the state of matter produced.

The Solenoidal Tracker At RHIC (STAR) detector is a large acceptance
spectrometer primarily designed to study hadron production in nuclear
collisions ranging from proton-proton to central Au+Au.
During the first year of operation, the STAR detector comprised a central
Time Projection Chamber (TPC), a Ring Imaging Cherenkov (RICH), a Central
Trigger Barrel (CTB) and two Zero Degree Calorimeters (ZDC). The TPC 
operated in a 0.25 Tesla solenoidal magnet and gives tracking 
information for charged particles in a pseudorapidity
interval of approximately $|\eta|<1.8$. It also provides a means of
particle identification from ionisation (energy loss) measurements along
each particle trajectory. Particle identification at higher
momenta was made possible using the RICH. The RICH covers only a
narrow acceptance window centered at mid-rapidity ($\Delta\phi =
20^{\circ}$ in azimuth and $|\eta|<0.3$), thus its measurements in year-1
were statistics limited. Overall about 500,000 Au+Au collisions 
(central and minimum bias) recorded with the STAR detector in year-1 
were analyzed for the results shown here. The detector was triggered using 
the signals from the CTB and ZDCs. 

\section{Results}

\begin{figure}
\begin{minipage}[h]{0.45\textwidth}
 \vspace{0.4cm}
 \includegraphics[width=\textwidth]{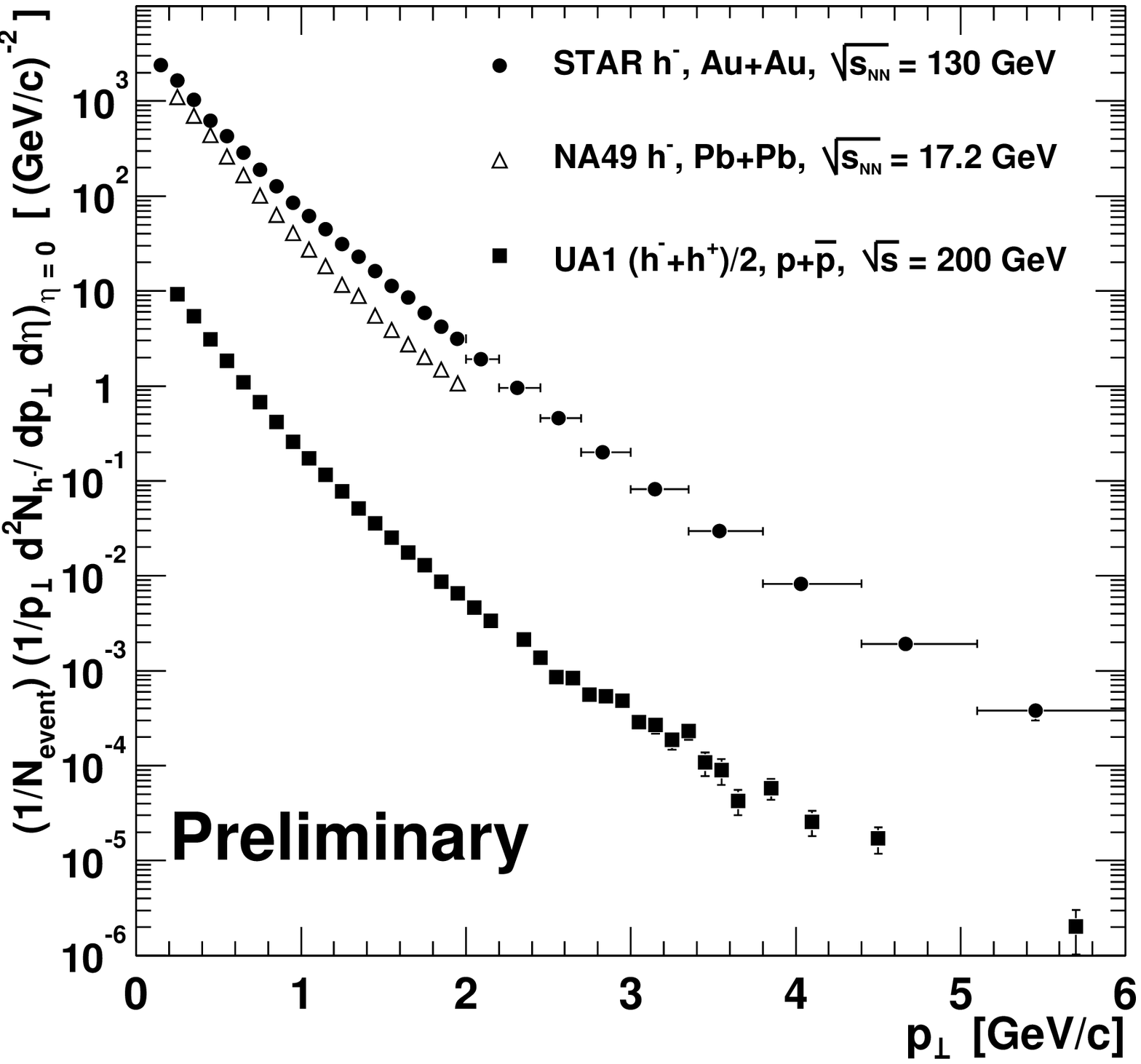}
 \vspace{0.1cm}
 \caption{The negative hadron $p_T$ distribution from the 5\% most central
 collisions. Also shown is the same distribution measured in Pb+Pb collisions
 at $\sqrt{s_{_{NN}}} = 17.2$~GeV and average charge hadron yield from 
 $p+\overline{p}$ collisions at $\sqrt{s} = 200$~GeV.}
 \label{pthminus}
\end{minipage}\hfill
\begin{minipage}[h]{0.45\textwidth}
 \vspace{0.4cm}
 \includegraphics[width=\textwidth]{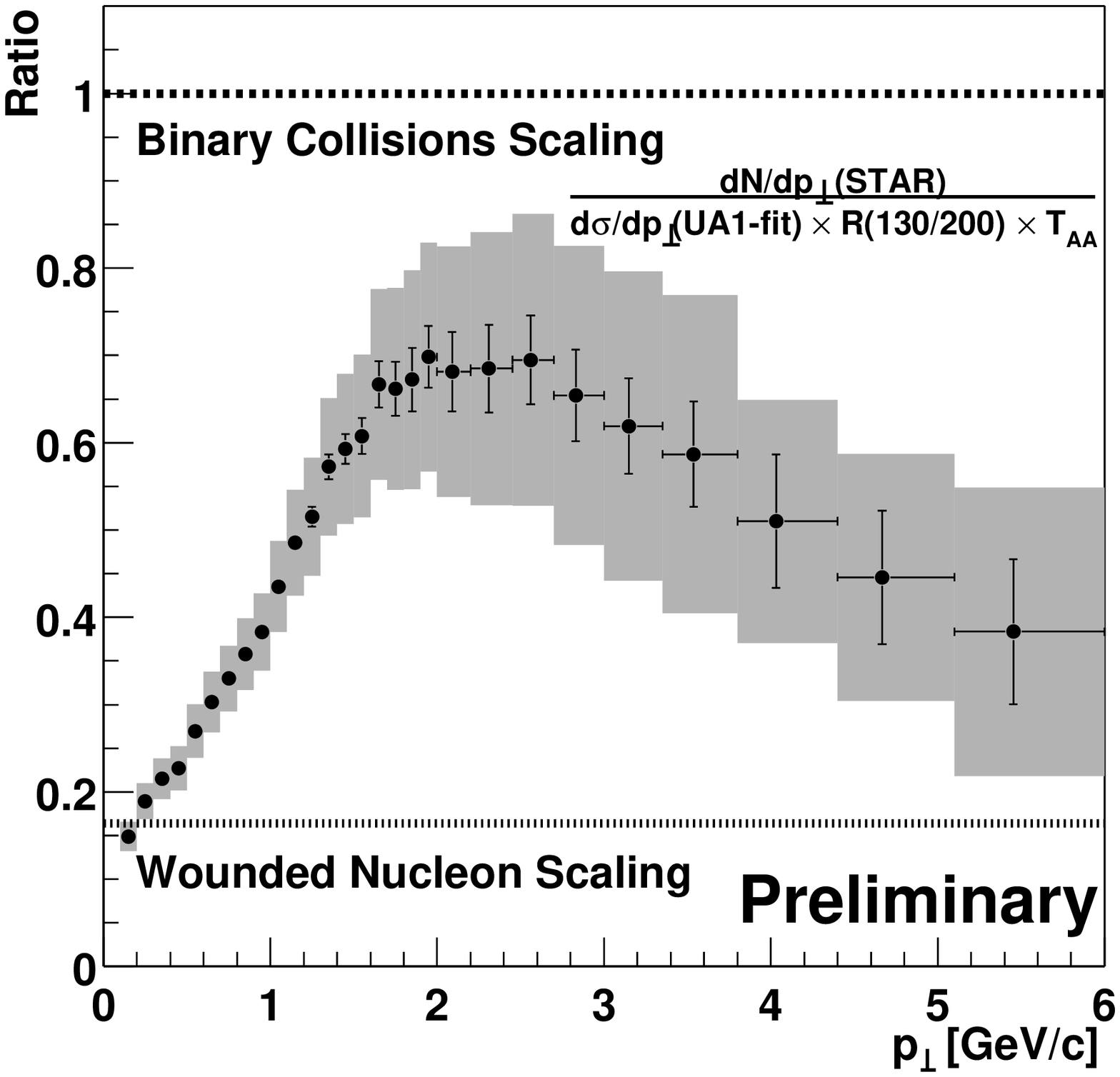}
 \vspace{0.1cm}
 \caption{Ratio of the STAR and scaled UA1 $p_T$ spectra. The vertical
          error bars indicate the measurement error, the shaded region
          the total uncertainty including that of the scaling factors
          applied to the UA1 data.}
 \label{ptratio}
\end{minipage}
\end{figure}

Fig.~\ref{pthminus} shows the inclusive transverse momentum distribution
of negative hadrons measured out to $p_T = 6$ GeV/$c$. The data shown are
for the 5\% most central event. The data has been corrected for background
arising from interactions in the detector and the products of weak decay
processes, as well as for losses due to the reconstruction efficiency. 

\begin{figure} 
\begin{minipage}[h]{0.45\textwidth}
 \includegraphics[width=\textwidth]{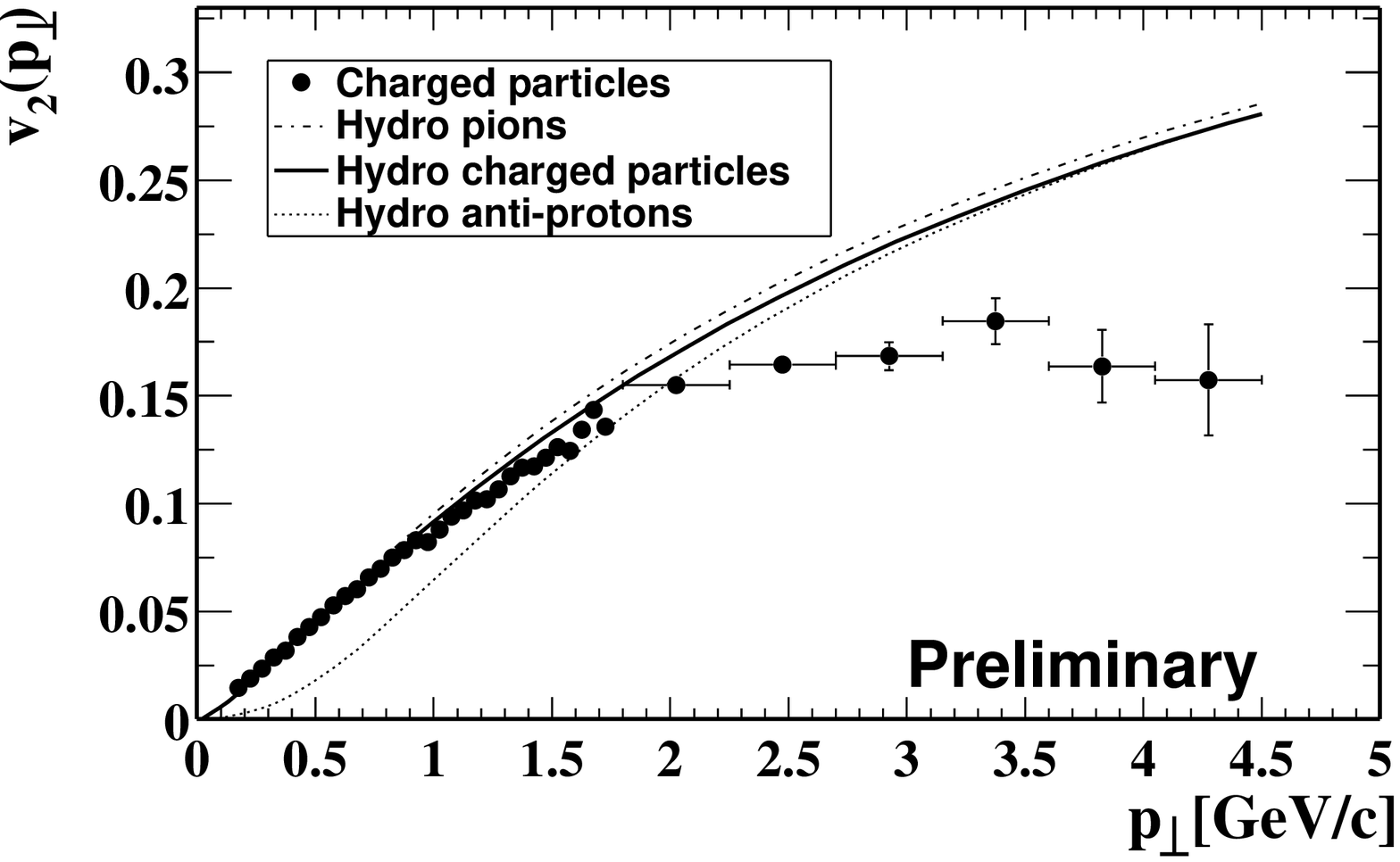} 
 \vspace{0.1cm}
 \caption{$v_2$ as function of $p_T$, compared to hydrodynamical calculations
          (see text for details).} 
 \label{v2hminus}
\end{minipage}\hfill
\begin{minipage}[h]{0.45\textwidth}
 \includegraphics[width=\textwidth]{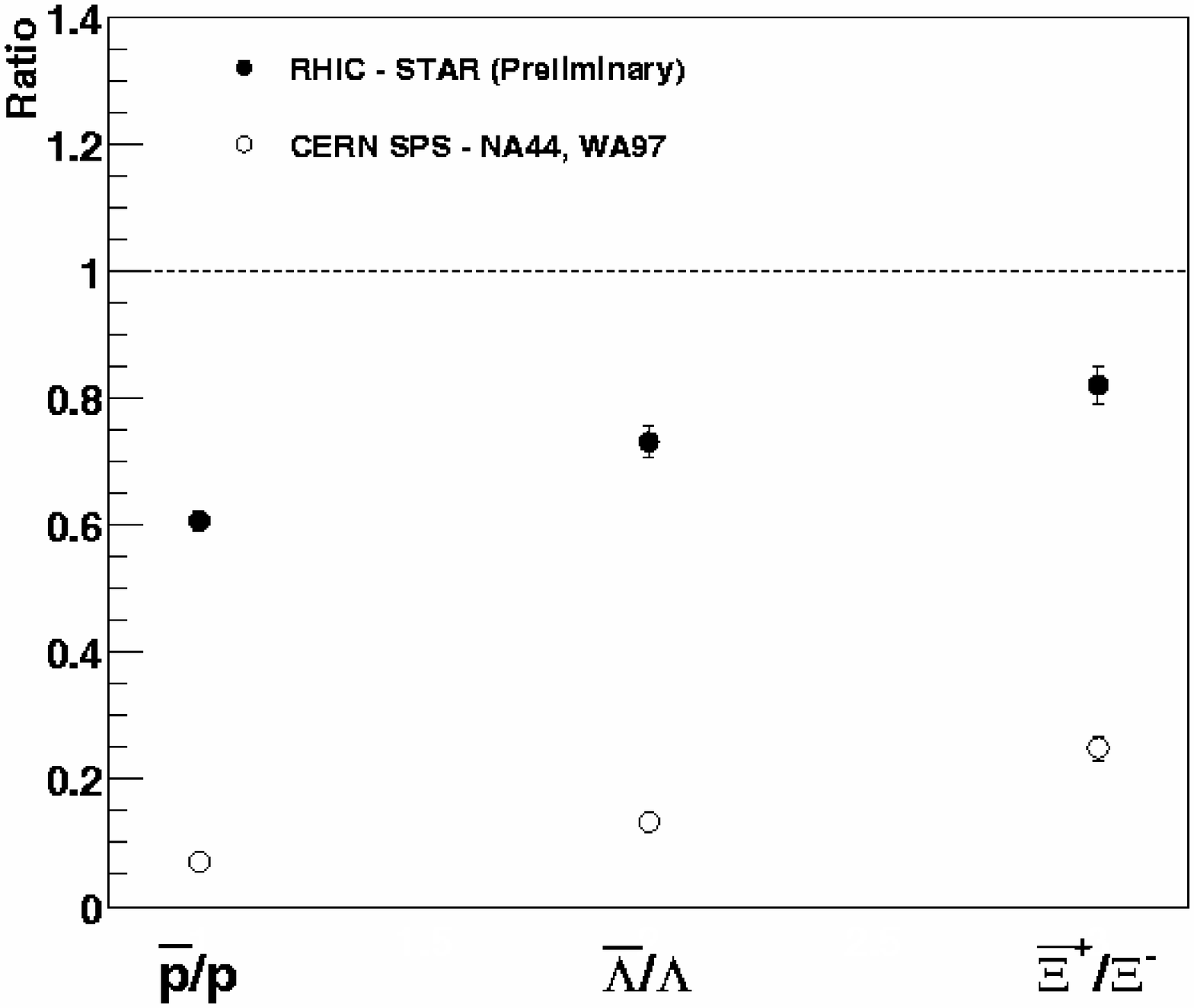} 
 \vspace{0.1cm}
 \caption{Anti-baryon to baryon ratios measured by STAR, ordered by strangeness
	  content and compared to values obtained at the SPS.}
 \label{ratios}
\end{minipage}
\end{figure}

Also shown in Fig.~\ref{pthminus} is the negative hadron distribution
from Pb+Pb collisions at $\sqrt{s_{_{NN}}} = 17.2$~GeV of
NA49~\cite{NA49hm} and the distribution of charged hadrons $(h^-+h^+)/2$
from proton-antiproton collisions at $\sqrt{s} = 200$~GeV of
UA1~\cite{UA1hpm}. It can  be seen that the STAR data is flatter than that
measured in Pb+Pb collisions at lower energy, corresponding to an increase
in the mean $p_T$ of negative hadrons by 18\%. The total yield, which was
obtained by extrapolation of a power-law fit to the data~\cite{STARhminus},
was found to be $dN/d\eta|_{\eta=0} = 280 \pm 1(stat) \pm 20(syst)$, a
52\% increase compared to the NA49 measurement. In order to compare to the
proton-antiproton data, Fig.~\ref{ptratio} shows the ratio of the STAR
data to a scaled parametrisation of the UA1 data. The UA1 data has been
scaled to account for the energy difference in the nucleon-nucleon
centre-of-mass frame and by the nuclear overlap 
integral, T$_{AA}$ = 26 mb$^{-1}$,~\cite{STARhminus} to arrive at the yield 
appropriate for the number
of binary nucleon-nucleon collisions expected in central Au+Au collisions.
The horizontal lines shown in Fig.~\ref{ptratio} correspond to the expected
value of the ratio under two scaling assumptions. The upper line shows the
expected value if the distribution scales with the number of binary collisions. 
This line corresponds to a ratio of unity due to the scaling of the UA1 data.
The lower line corresponds to the expected value of the ratio if the
distributions scale with the number of wounded nucleons. 
As expected, at low $p_T$, where particle production is dominated by 
soft processes, the total pion yield scales with the number of wounded
nucleons. However, this is only the case for the lowest transverse
momentum bin. In general, the data below $p_T = 2$~GeV/$c$ exceeds
the wounded nucleon scaling, rising as a function of $p_T$ toward the
binary collision limit. This upper limit is relevant for hard processes
in the absence of any nuclear effects, such as initial state multiple scattering, 
jet quenching or collective radial flow. In fact, the ratio never reaches 
the binary collision scaling, but drops again for $p_T > 2$~GeV/$c$. At a
$p_T$ of 6~GeV/$c$, the ratio falls below the binary collision scaling by
more than a factor of two.
A second measurement that may be sensitive to high $p_T$ phenomena is
elliptic flow. Elliptic flow is studied in minimum bias collisions,
where the initial spatial anisotropy caused by the partial overlap of two
nuclei is transformed into a momentum anisotropy of particles in the final
state. For each event a reaction plane is determined from the predominantly
low $p_T$ tracks reconstructed in the TPC~\cite{STARflow}.
Particle distributions are
then studied for an azimuthal asymmetry with respect to the reaction plane,
which may stem from pressure gradients generated in the earliest stages of
the collision. Fig.~\ref{v2hminus} shows the second
Fourier coefficient, $v_2$, of the azimuthal distribution of negative
hadrons as a function of $p_T$. The data shows that $v_2$ rises as a
function of $p_T$, as expected from hydrodynamic calculations~\cite{Hydro},
until around $p_T = 2$~GeV/$c$ where-after the data falls systematically below
the hydrodynamical prediction. This departure is not unexpected as high $p_T$
particles may escape the reaction zone with little rescattering. However,
$v_2$ appears to saturate rather than decrease and this may be related to
the in-medium energy loss of high $p_T$ partons.
Calculations have shown that in this case the magnitude
of $v_2$ is sensitive to the initial gluon density~\cite{Quench}.

It is tempting to connect the two observations seen in Figs.~\ref{pthminus}
and~\ref{v2hminus}. In Fig.~\ref{pthminus} a suppression of high $p_T$ 
hadrons is observed in Au+Au collisions relative to scaled $p+\overline{p}$
data, while in Fig.~\ref{v2hminus} the anisotropic flow signal is found to
saturate at high $p_T$, both of which may be the result of partonic energy
loss (jet quenching) in the excited medium.

Particle identification in STAR can be performed in three ways: using
the specific ionisation ($dE/dx$) in the TPC, from the angle at which
Cherenkov light is emitted from particles entering the RICH, and by
the topology of weak decays reconstructed from the tracks in the TPC.
At a given momentum, the difference
in the mean ionisation between particle species allows the separation of pions
and kaons for momenta $p \leq 0.6$~GeV/$c$ and the separation of kaons and
protons for momenta $p \leq 1.0$~GeV/$c$. The RICH extends the particle
identification capability of the TPC to around 3~GeV/$c$ for kaons and 
5~GeV/$c$ for protons. The topological pattern recognition technique is
applicable to a wide range of weak decay processes, from neutral strange
particle decays such
as $K^0_s \rightarrow \pi^+ + \pi^-$ and $\Lambda \rightarrow p + \pi^-$,
to cascade decay processes such as $\Xi^- \rightarrow \Lambda + \pi^-$ and
one-prong decays such as $K^- \rightarrow \mu^- + \nu_\mu$. 

\begin{figure} 
\begin{minipage}[]{0.5\textwidth}
 \vspace{-0.5cm}
 \includegraphics[width=\textwidth]{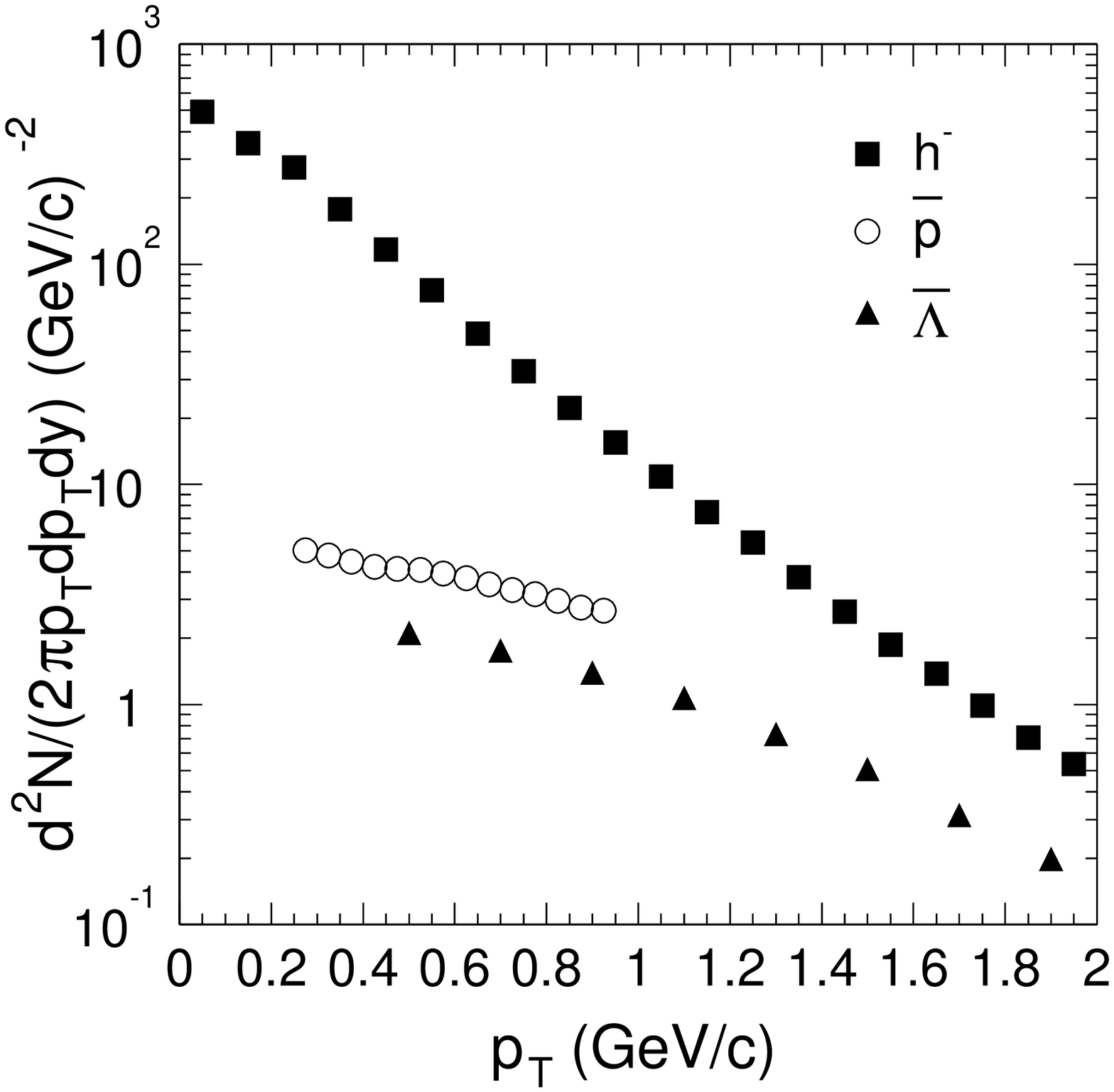} 
 \caption{The transverse mass spectra of $h^-$, $\overline{p}$
          and $\overline{\Lambda}$ in the 11\% most central events.}
 \label{HydroNew}
 \centering{\vspace{-6.4cm} Preliminary\vspace{6.4cm}}
\end{minipage}
\begin{minipage}[]{0.5\textwidth}
 \includegraphics[width=\textwidth]{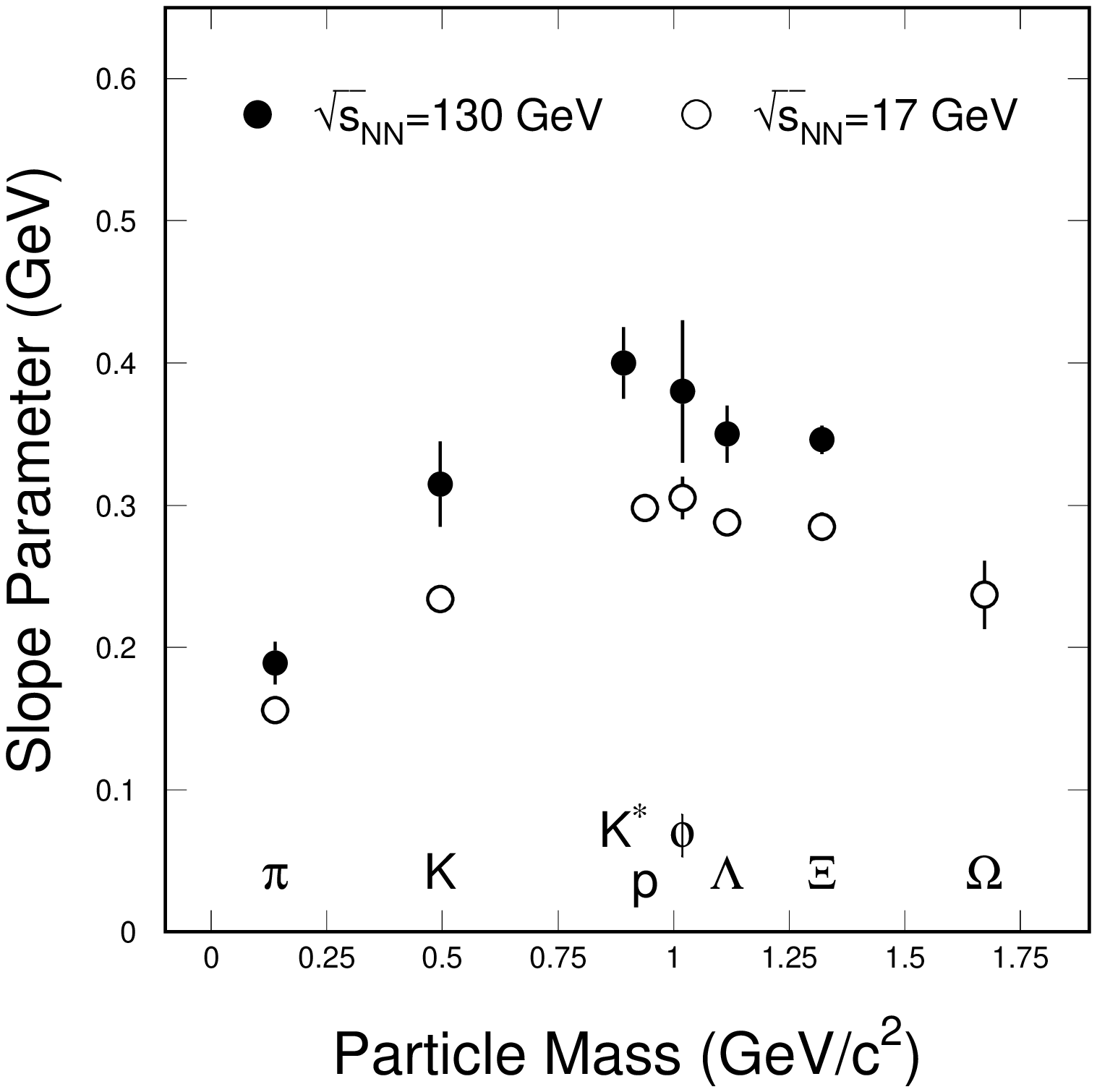}
 \vspace{-1.cm}
 \caption{The mass dependence of the slope parameters obtained from exponential
          fits to the transverse mass distribution of strange particles 
          in the central Au-Au collisions (statistical errors only).} 
 \label{allslopes}
 \centering{\vspace{-6.2cm} Preliminary\vspace{6.2cm}}
\end{minipage}
\end{figure}

Fig.~\ref{ratios} shows the anti-particle over particle ratios for identified
baryons at mid-rapidity in the 11\% most central collisions. The ratios 
indicate that the mid-rapidity region is not net baryon free, although 
comparisons with measurements at SPS energies indicate that the net baryon
density is much reduced. The ratios are consistent with simple
quark counting models ~\cite{quark} and statistical
thermal models which are governed by a common baryon chemical
potential and chemical freeze-out temperature ~\cite{pbm}. The ratios
shown in Fig.~\ref{ratios} are based on integrated yields, which are
easy to obtain in the case of the strange baryons, where the topological
pattern recognition allows to cover around 70\% of the particle's
momentum spectrum. In the case of the $\overline{p}/p$ ratio the
coverage based on the energy loss technique is much smaller due to the 
rather low momentum cutoff ($p \leq 1$~GeV/$c$). Still, an extrapolation was
done and the details can be found at ~\cite{STARpbarp}. 
The particle identified transverse mass distributions of negative hadrons
($\pi^-$, K$^-$ and $\overline{p}$) obtained by the $dE/dx$ method 
in the 6\% most central collisions were compared to an exponential fit. 
A clear mass dependence was observed in the inverse slope parameter,
with values ranging from 190~MeV for $\pi^-$ to 300~MeV for K$^-$ and 
565~MeV for $\overline{p}$. These observations are suggestive of a 
significant increase in transverse radial flow compared to CERN 
observations~\cite{NA49hm,CERNFlow}. Unfortunately the transverse mass range 
obtained from the $dE/dx$ measurement is very limited, especially for 
the $\overline{p}$. The transverse momentum spectra of $h^-$,
$\overline{p}$ and $\overline{\Lambda}$ are shown in Fig.~\ref{HydroNew}.

It is noteworthy that the inverse
slope obtained from an exponential fit to the $\overline{\Lambda}$ spectrum 
is significantly lower than that of the $\overline{p}$ spectrum. One of the 
consequences of radial flow is a flattening of the transverse momentum 
spectra for heavier particles at low $p_T$~\cite{Rasmussen}. Therefore 
the slope obtained from a  simple exponential fit to the data will be 
dependent on the range of $p_T$ used in the fit. Fig.~\ref{HydroNew} also 
suggests that the relative yield of $\overline{\Lambda}$ to negative hadrons 
becomes increasingly larger at higher $p_T$. It should be noted that the 
negative hadron distribution already contains the yield of primary antiprotons.
If this trend continues, then it would appear that the yield of antibaryons
may exceed that of mesons at transverse momenta above 2~GeV/$c$. This is
in qualitative agreement with the direct measurement of the $\overline{p}/\pi^-$
reported by the PHENIX collaboration~\cite{PHENIX}. A detailed study of 
baryon production in year-2 may determine whether this observation is solely a 
consequence of transverse radial flow, or due to novel baryon 
dynamics~\cite{Vitev}.

In an attempt to remove the dependence of the fit range and seek a common
explanation for the spectra, a hydrodynamical motivated fit~\cite{Schnedermann}
has been performed. The model allows a choice of the transverse flow profile
of matter produced in the collision. The choice which best describes the
data is a radial flow velocity which varies as $\beta_r = \beta_s\sqrt{r/R}$ 
where $\beta_s$ is the surface flow velocity at a radius, $R$. A fit was
performed to each of the particle spectra, allowing the freezeout temperature
and mean flow velocity to vary. It was found that a freezeout temperature 
of $T = 120^{+50}_{-25}$~MeV and an average flow velocity 
of $\beta_r = 0.52^{+0.12}_{-0.08}$ could simultaneously describe all
measured particle spectra. These values represent a modest rise in the 
transverse flow velocity and similar freezeout temperature compared to 
earlier measurements at CERN~\cite{NA49hm,CERNFlow}.
For the strange particles reconstructed in STAR (including the Kaon kink
analysis, which extends the m$_{T}$ coverage up to 2 GeV/c) the coverages
are similar and a simple exponential fit can be applied in order to yield
a common mass dependence. The result is shown and compared to SPS
data in Fig.~\ref{allslopes}.

\section{Summary}

In summary, STAR has measured a wide range of hadronic probes in both
central and minimum bias Au+Au collisions at $\sqrt{s_{_{NN}}} = 130$~GeV.
The pseudorapidity density of negative hadrons
is $dN/d\eta = 280 \pm 1 \pm 20$ in central collisions, representing a 
52\% increase compared to lower energy data from CERN. The elliptic
flow measurement in minimum bias collisions is consistent with the
predictions of a hydrodynamical model at low transverse momentum, 
implying that the system equilibrates early. Studies of the transverse
momentum spectra of identified particles produced in central collisions
indicate the presence of a strong collective radial flow, which
complicates the interpretation of transverse momentum spectra. However,
a common description of the spectra is possible within a hydrodynamical
framework. At transverse momenta greater than 2~GeV/$c$ there is a 
suppression of negative hadrons compared to an extrapolation from 
proton-antiproton collisions. The elliptic flow measurement is also found 
to saturate in this momentum region. Both observations may be sensitive to 
partonic energy loss (jet quenching) at this higher energy. Finally it has 
been shown that antibaryons (and baryons) become an increasingly larger 
fraction of the total particle yield at high $p_T$.

\end{document}